\documentclass[11pt]{article}
\usepackage{iap2000,epsfig}

\def\etal{\em et al.}

\def\be{\begin{equation}}
\def\ee{\end{equation}}
\def\bea{\begin{eqnarray}}
\def\eea{\end{eqnarray}}

\begin{document}
\title{THE SUNYAEV-ZEL'DOVICH EFFECT WITH THE COSMIC BACKGROUND IMAGER}

\author{P.S. Udomprasert, B.S. Mason, and A.C.S. Readhead}

\address{Department of Astronomy, 105-24 Caltech, Pasadena, CA, 91125}

\maketitle

\abstracts{ We are engaged in a program with the Cosmic Background
Imager (CBI), a 13-element interferometer recently installed at 5000 m
in the Chilean Andes, to measure $H_0$ with 10\% accuracy through
observations of the Sunyaev-Zel'dovich effect in a sample of nearby
clusters.  We discuss the capabilities of this new instrument and
present our cluster sample and selection criteria, along with results
from detailed simulations.  We also present preliminary observations
of Abell 478.}

\section{The Cosmic Background Imager}

\begin{figure}
\begin{center}
\psfig{figure=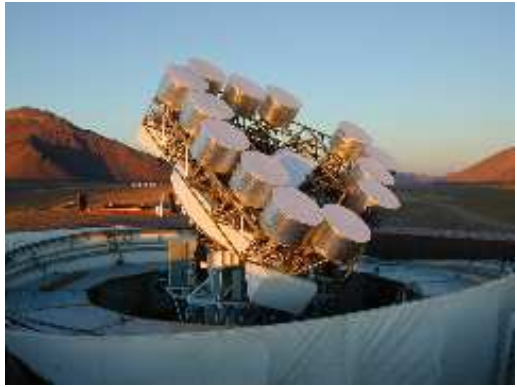,height=2in}
\end{center}
\caption{The CBI, installed at 5000 m in the Chilean Andes}
\label{fig:cbiphoto}
\end{figure}

The Cosmic Background Imager (CBI), shown in Fig.~\ref{fig:cbiphoto},
is a radio interferometer array whose design makes it particularly
suitable for observations of the SZE in low-$z$ clusters ($z<0.1$).
It can therefore be used to determine the value of $H_0$ with high
accuracy and provide us with insight into the physical state of the
cluster gas.

\subsection{Instrument Specifications}\label{subsec:specs}

The CBI is a 13-element interferometer mounted on a 6 meter platform
operating in ten 1-GHz frequency bands from 26 GHz to 36 GHz.  Dishes
90 cm in diameter accommodate a range of baselines from 100 cm to 550
cm.  The instantaneous field of view of the instrument is $44'$ and its
resolution ranges from $3'$ to $10'$, depending on configuration.  HEMT
amplifiers cooled to 6 K and low atmospheric noise at the high, dry
site, allow a system temperature of $\sim$30 K.  In 5 hours, the CBI
can produce images with an rms noise per beam of 1.4 mJy, which
corresponds to 30 $\mu$K for a synthesized beam of $4.5'$ FWHM.  Typical
SZ decrements at 31 GHz are of order several hundred $\mu$K.

\subsection{SZ with the CBI}

\begin{figure}
\begin{center}
\psfig{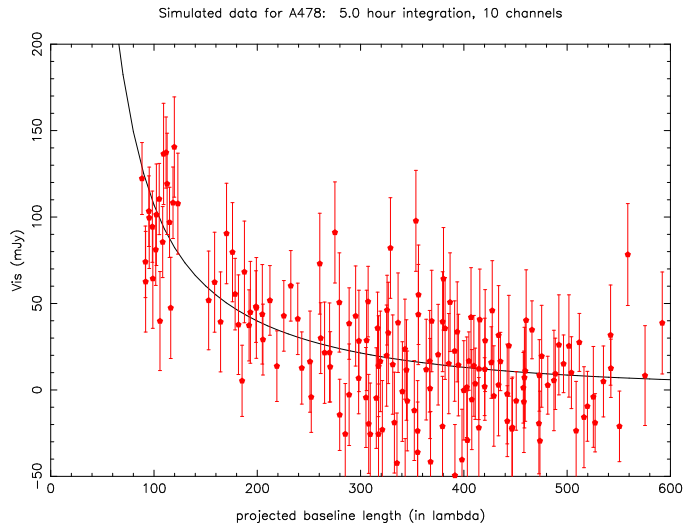}
\end{center}
\caption{Model visibility profile for A478 with simulated CBI data}
\label{fig:radpl}
\end{figure}

The CBI's angular resolution makes it well-suited for observations of
low-redshift clusters.  Fig.~\ref{fig:radpl} shows a simulated CBI
visibility data set for A478, a cluster at $z=0.0881$\cite{SR}.  The
model profile was calculated using an isothermal $\beta$-model with
cluster parameters $\beta$=0.638, core radius $\theta_0$=$1'$, and
central gas density $n_0 = 0.028 h_{50}^{1/2} \mbox{cm}^{-3}$, derived
from ROSAT X-ray data by Mason and Myers \cite{MM}.  We assumed
spherical symmetry and used a gas temperature $T_e$ = 8.4 keV
\cite{ASCA}.  The visibility profile demonstrates the CBI's special
suitability for imaging the SZE in low-redshift clusters.  For a large
extended source, the interferometer's response drops rapidly for
baselines between 100 and 200 wavelengths, which at 30 GHz corresponds
to a baseline length of 1 to 2 meters.  Much larger interferometers
observing at the same frequencies aren't sensitive to this region of
u-v space and must observe more distant clusters ($z>0.15$) which have
smaller angular sizes and are therefore better matched to the response
of their longer baselines.  The CBI's exceptional sensitivity allows
us to observe a large sample of nearby clusters.  Each cluster only
requires about 2-3 nights of observing time.

We combine these advantages in our program to observe the SZE in a
complete sample of 19 low-$z$ galaxy clusters.  In doing so we hope to
address three key issues that can lead to inaccurate determinations of
$H_0$: cluster asphericity, clumpy gas distribution, and
non-isothermal gas.  Observation of a large, complete sample should
eliminate any bias in $H_0$ due to elongation effects, while the focus
on nearby clusters will allow us better to understand the effects of a
clumpy gas distribution.  When combined with the spectral imaging
capabilities of XMM-Newton and Chandra this reveals the temperature
structure of galaxy clusters, which are important factors in modeling
the expected SZE.

\section{The Distance Scale and SZ}

Accurate knowledge of the Hubble Constant allows us to calculate
fundamental cosmological parameters such as the age and size scale of
the universe, as well as important astrophysical quantities such as
luminosities and physical sizes.  While groups such as the HST $H_0$
Key Project are measuring $H_0$ to high precision \cite{HST}, it is
important to obtain a measurement of comparable accuracy via
completely independent methods.

The Sunyaev-Zel'dovich effect is caused by the inverse Compton
scattering of Cosmic Microwave Background (CMB) photons off electrons
in a hot ionized gas.  The scattering boosts the low energy photons to
higher energies, so we see a distortion in the CMB spectrum, a
decrement in intensity at low frequencies and an increment at high
frequencies \cite{SZ}.  Coupled with X-ray observations, the SZE
provides a direct measurement of $H_0$, independent of the astronomical
distance ladder.  The SZE is proportional to $\int n_eT_e dl$, while
the X-ray emission due to thermal bremsstrahlung is proportional to
$\int n_e^2\Lambda_{eH} dl$, where $n_e$ is the electron density,
$T_e$ is the electron temperature, and $\Lambda_{eH}$ is the X-ray
cooling function, which depends on temperature.  The X-ray
observations constrain the density and temperature profiles, allowing
one to predict the expected SZE towards a cluster.  The comparison of
the X-ray and SZE observations, coupled with the assumption that
clusters are spherically symmetric, provides a measurement of $H_0$.

Accurate detections of the SZE are now becoming fairly routine.  See,
for example, Carlstrom {\etal}\cite{JC}, Grainge {\etal}\cite{Ryle},
Myers {\etal}\cite{STM}, and Holzapfel {\etal}\cite{Suzie}.  However,
other than Myers {\etal}, most groups have focused on high redshift
clusters where it is more difficult to define a complete sample.
Also, as mentioned above, there are still a number of difficulties
associated with the SZE/X-ray determination of $H_0$:

\begin{enumerate}
\item{Cluster asphericity:}

The observational evidence suggests clusters are not spherically
symmetric \cite{HB}.  One can overcome this by selecting a complete
sample of clusters and averaging the $H_0$ measurements over the
entire sample.  The random orientations will cancel out statistically,
providing one with an unbiased measurement of $H_0$.  We plan to
accomplish this using an X-ray flux-limited sample, discussed below.

\item{Sub-clumping of cluster gas:}

The cluster gas is usually represented by a smooth $\beta$-model profile.
Other SZ studies don't take into account the effect of a clumpy gas
distribution, and we hope to do this by observing nearby clusters,
where the density structure will be easier to observe and understand.

\item{Non-isothermal cluster gas:}

The value of $H_0$ depends strongly on the model used for the gas
temperature profile.  Cluster temperature profiles have been difficult
to constrain due to insufficient X-ray data, but detailed XMM
observations should remedy this problem.
\end{enumerate}

\section{Sample Selection}

We wish to minimize the effects of cluster asphericity by studying a
complete orientation unbiased sample.  To compile such a sample, we
combined the results of ROSAT cluster surveys by Ebeling {\etal}
1996\cite{E2}, 1998\cite{E1}, and De Grandi {\etal} \cite{DG} and
imposed a flux limit $f_{0.1-2.4 \mbox{ keV}} > 1.0 \times 10^{-11}$
erg cm$^{-2}$ s$^{-1}$ which is significantly higher than the expected
completeness levels of the various catalogs.  We then assembled a
volume complete sample out to our redshift cutoff $z<0.1$ by only
including clusters with $L_{0.1-2.4 \mbox{ keV}} > 4.52 \times 10^{44}
h_{50}^{-2}$ erg s$^{-1}$.  This luminosity cutoff should also
eliminate any possible optical bias in the catalogs by removing the
poorer clusters from the sample, as discussed in detail by Mason and
Myers\cite{MM}.  Our sample contains 25 clusters which are observable
with the CBI.  The 19 most luminous clusters have pointed ROSAT and
ASCA observations available, and these clusters constitute our primary
sample.

\section{CBI Observations: Abell 478}

\begin{figure}
\begin{center}
\psfig{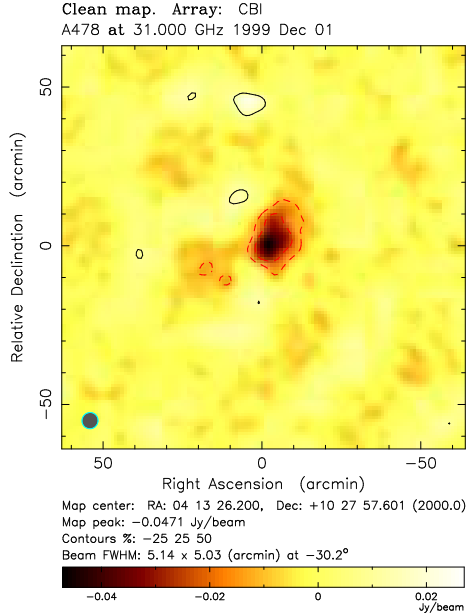}
\end{center}
\caption{A CBI image of A478}
\label{fig:a478}
\end{figure}

We observed Abell 478 with a subset of the CBI (9 of 13 antennas and 7
of 10 channels) in December 1999.  To remove the ground spillover
detected at the $\sim$ 1 Jy level on short (100 cm) baselines, we
observed lead and trail fields at the same declination and hour angle
range, and subtracted the average of the lead and trail fields from
the cluster field.  This method allows for accurate removal of the
ground signal, but increases the rms noise by a factor of $\sqrt{1.5}$
and triples the on source integration time.  Fig.~\ref{fig:a478} shows
a CBI image of A478, taken with the partial array in December over
about 6 hours (on source time).  The noise is 4.2 mJy/beam, so the
peak detection is about 11-$\sigma$.  The CBI has been in full
operation since January 2000.  We have preliminary observations of
A85, A1651, A2029, A2384, A2597, A3158, A3667, and A3921, and we will
continue to conduct observations of these and the remaining clusters
during the next year.

\section{Tentative Error Budget}

Intrinsic anisotropy fluctuations are the largest source of random
error in the determination of $H_0$ from low-$z$ clusters.  Our
simulations indicate that errors in $H_0$ of 35\% to 40\% per cluster
due to intrinsic anisotropy would be typical.  The expected random
error in the determination of $H_0$ from sources including these
fluctuations, asphericity, residuals from point source subtraction,
observational noise and calibration, and derivation of cluster
parameters from X-ray observations is about 45\%.  For a sample of 19
clusters, we expect a random error of about 10\% in our final
determination of $H_0$.

By accounting for clumpy, non-isothermal gas distributions we should
minimize the systematic errors from these sources, although we
currently cannot quantify by how much.  We anticipate a systematic
error due to CBI absolute flux calibration of about 5-6\%.

\section*{Acknowledgments}
The CBI is a project of the California Institute of Technology, in
collaboration with the University of Chile.  The construction of the
CBI has been made possible by the generous support of the California
Institute of Technology, Ronald and Maxine Linde, Cecil and Sally
Drinkward, and grants from the National Science Foundation (awards
AST-9413935 and AST-9802989).  PSU gratefully acknowledges support
from the NSF Graduate Fellowship program.

\end{document}